\documentstyle[11pt,aaspp4]{article}

\slugcomment{Submitted to The Astronomical Journal}

\lefthead{Kirkpatrick et al.}
\righthead{Brown Dwarf Companions}

\begin{document}

\title{Brown Dwarf Companions to G-type Stars. I: Gliese 417B and Gliese
584C\footnote{Portions
of the data presented herein were obtained at the W.M. Keck Observatory which
is operated as a scientific partnership among the California Institute of
Technology, the University of California, and the National Aeronautics and
Space Administration.  The Observatory was made possible by the generous
financial support of the W.M. Keck Foundation.}} 

\author{J. Davy Kirkpatrick}
\affil{Infrared Processing and Analysis Center, MS 100-22, California 
    Institute of Technology, Pasadena, CA 91125; davy@ipac.caltech.edu}

\author{Conard C. Dahn, David G. Monet}
\affil{U.S.\ Naval Observatory, P.O. Box 1149, Flagstaff, AZ 86002; 
dahn@nofs.navy.mil, dgm@nofs.navy.mil}

\author{I. Neill Reid}
\affil{Space Telescope Science Institute, 3700 San Martin Drive, Baltimore, 
MD 21218; inr@stsci.edu; also Department of Physics and Astronomy, University 
of Pennsylvania, Philadelphia, PA 19104-6396}

\author{John E. Gizis}
\affil{Infrared Processing and Analysis Center, MS 100-22, California 
    Institute of Technology, Pasadena, CA 91125; gizis@ipac.caltech.edu}

\author{James Liebert}
\affil{Steward Observatory, University of Arizona, Tucson, AZ 85721;
liebert@as.arizona.edu}

\and

\author{Adam J. Burgasser}
\affil{Department of Physics, MS 103-33, California 
    Institute of Technology, Pasadena, CA 91125; diver@its.caltech.edu}

 

\begin{abstract}

We present astrometric and spectroscopic observations confirming that two
nearby G dwarf systems (Gliese 417 = BD+36$^{\circ}$2162 and Gliese 584AB =
$\eta$ CrB AB)
have a widely separated, L dwarf, substellar companion. 
Using age estimates of the G dwarf primaries, we estimate masses for these 
L dwarfs from theoretical evolutionary tracks.  For the L4.5 dwarf Gl 417B we
estimate an age of 0.08-0.3 Gyr and a mass of 0.035$\pm$0.015 $M_\odot$.
For the L8 dwarf Gl 584C we estimate an age of 1.0-2.5 Gyr and 
a mass of 0.060$\pm$0.015 $M_\odot$.  This latter object also shows
evidence of spectrum variability, which may be 
due to surface inhomogeneities rotating into and out of view.  These
new companions are also compared to six other
L dwarf and T dwarf companions previously recognized.  For the L dwarf 
companions, ages implied by the presence or absence
of lithium are consistent with ages inferred from the primaries
alone.

\end{abstract}




%

\section{Introduction}

The search for brown dwarfs and planets as close companions to nearby stars has
met with spectacular success in recent years.  Radial
velocity surveys of F, G, K, and M stars have turned up many planetary
companions with
masses of $\sim$0.001 M$_\odot$ and higher (see Marcy \& Butler 2000 and 
references therein).  High spatial resolution space-based studies have also
revealed low-mass stellar and substellar
companions to nearby stars and brown dwarfs (HST NICMOS --
Lowrance et al.\ 1999, Lowrance et al. 2000, Mart{\'{\i}}n et al.\ 1999; HST
WFPC2 -- Reid et al.\ 2000).  Ground-based adaptive optics, coronagraphy, 
and small-field imaging have also resulted in several important discoveries
(e.g., Nakajima et al.\ 1995, Rebolo et al.\ 1998).

Despite the recent successes, the search for similar companions at much wider 
separations has seen little advancement since the days of van Biesbroeck's 
(1944, 1961) photographic
search for common-proper-motion companions to high motion
stars.  Finding companions at wide 
separations involves one main limitation: large pieces of sky have to
be canvassed.  In the case of the triple
star $\alpha$ Centauri AB + Proxima Centauri, the apparent separation is a
staggering 2.18 degrees, which at the system's distance of 1.3 pc corresponds
to a separation of $\sim$0.04 pc.  Even wider physical 
separations are known, however, such as the $\gamma$ Ceti AB + Gl 
106.1C system with a separation of $\sim$0.09 pc.  

The best available observational 
evidence suggests that widely separated companions exist out 
to distances of $\sim$0.1 pc (Latham et al. 1984, Weinberg 1990).  Companions 
at significantly wider separations are highly unlikely due to dynamical 
interactions with passing field stars (Retterer \& King 1982).  Using data from
Close, Richer, \& Crabtree (1990) it is found that $\sim$5\% of systems 
(or $\sim$3\% of stars) with
$M_V \le 9.0$ (spectral type $\sim$M0 V) have a companion wider than 0.01 pc.
When M dwarfs are also included in the analysis, it is found that a similar
number ($\sim$6\% of systems) have companions at these separations
(Fischer \& Marcy 1992).

Estimates by Reid et al.\ (1999a,b) show that there could be perhaps as many as
two times the number of brown dwarfs as stars 
in the solar neighborhood.  If the 
distribution of substellar companions mimics that of stellar companions
we might expect up to 10\% of disk stars to have a brown dwarf companion
at a separation greater than 0.01 pc.  However, the results of Marcy \& Butler
(2000) show that there is a distinct difference between the frequency of
stellar and substellar companions at very close separations.  
They note a paucity of brown dwarf companions -- the ``brown dwarf desert'' --
at separations less than 3 AU around solar-type main sequence stars; fewer 
than 0.5\% of their stars have a
brown dwarf companion at these separations.  The reason for this difference
remains unexplained, but a possible key is in finding whether a similar
``brown dwarf desert'' exists at wider separations from main sequence stars.

Today's deep, large-area surveys like the Deep Near-Infrared Survey (DENIS; 
Epchtein et al.\ 1997), the Two Micron All Sky Survey (2MASS; Skrutskie et 
al.\ 1997), and the Sloan Digital Sky Survey (SDSS; York et al.\ 2000) are 
presenting a wealth of new material with which lower luminosity companions to 
nearby stars can be found at very wide apparent separations.  Two
such companions with K dwarf primaries have already been found by 2MASS and 
have been announced elsewhere -- Gl 570D (Burgasser et al.\ 2000) and GJ 1048B
(Gizis et al.\ 2000).  Here we discuss two additional brown 
dwarfs, Gl 417B and Gl 584C, both of which are L dwarf companions to nearby G 
dwarfs.

The discovery of these L dwarfs is described in \S2.  Details on
the spectroscopic follow-up are given in \S3 along with 
astrometric measurements showing that the two L dwarfs share common proper
motion with the nearby G stars.  Using extensive supplementary observations
of the
G dwarf primaries, we derive age estimates for the systems and mass estimates
for the L dwarf companions in \S4.  We compare these two brown dwarfs
to other known companions in \S5 and summarize our conclusions in \S6.

\section{Discovery}

While searching the 2MASS data for late-L dwarf suspects (Kirkpatrick et al.\
2000), we uncovered
two candidates lying in close proximity to known nearby stars.  The
first candidate has a designation of 2MASSI J1112256+354813 
and is shown in the finder chart of Figure 1a.  As an
L-dwarf candidate it is bright and quite red: $K_s = 12.69{\pm}0.05$ and 
$J-K_s = 1.88{\pm}0.06$.  As shown in the figure it
lies 90\arcsec\ southwest of the nearby G dwarf Gl 417 (aka HR 4345,
HD 97334, BD+36$^\circ$2162)\footnote{The field around Gl 417A contains several
relatively bright stars.  BD+36 2164 (V=7.2) lies $\sim$156\arcsec\ to the 
northeast of Gl 417A (V=6.4).  The Hipparcos parallax and
proper motion measurements (Perryman et al.\ 1997) for this star show that
it is unassociated with Gl 417A.  A dimmer star
(V=12.3) referred to as BD+36 2162C
is located $\sim$86\arcsec\ southeast of Gl 417A.  This star has no Hipparcos
measures, but it is also unassociated because a comparison of the 
DSS and XDSS images shows that it exhibits no detectable proper motion
(whereas that of Gl 417A is obvious).  Another bright star, BD+36 2165 (LHS
2373; V=9.8),
is located $\sim$312\arcsec southeast of Gl 417A.  This star exhibits motion 
that is larger and along a substantially different position angle than
that of Gl 417A, and Hipparcos also finds a much greater distance.  Hence, this 
object is also not associated.}. 
Hereafter we will refer to the primary as Gl 417A and to the L dwarf candidate 
as Gl 417B.

The second candidate has a designation
of 2MASSW J1523226+301456 and is shown in 
the finder chart of Figure 1b.  It is even redder than the previous
candidate -- $J-K_s = 2.09{\pm}0.13$ with $K_s = 14.24{\pm}0.07$ -- and 
lies 194\arcsec\ southeast of the nearby G dwarf double Gl 584AB 
(aka $\eta$ CrB AB, HR 5727/8, HD 137107/8, BD+30$^\circ$2653, 
ADS 9617AB)\footnote{Two other relatively bright stars are also listed in
various catalogues as members of this system.  A star known as BD+30$^\circ$
2653C (ADS 9617C, V=13.4) is located 72\arcsec\ due north of Gl 584AB.  
Another star, ADS 9617D (V=11.0) is located 217\arcsec\ to the northeast of
Gl 584AB.  A comparison of the 1954-epoch POSS-I images to the 2000-epoch
2MASS images shows that neither of these stars shares the proper motion of
Gl 584AB, so neither is physically associated.}.
Hereafter we will refer to the L dwarf candidate as Gl 584C.

As Figure 1 shows, neither of these L dwarf candidates is visible on the 
POSS-I E (red) plates, indicating very red optical-minus-infrared colors.  
The sources 
are also invisible in on-line images of the POSS-II J (blue) and F (red) 
plates.  Thus, the
2MASS images represent the first-epoch detections for both candidates.  

\section{Follow-up Observations}

\subsection{Astrometric Confirmation}

The G dwarf primaries have small motions on the sky -- 0{\farcs}29 yr$^{-1}$
for Gl 417A and 0{\farcs}22 yr$^{-1}$ for Gl 584C (Perryman et al. 1998) --
so moderately precise astrometry is needed for speedy verification of common
proper motion between the G dwarf primaries and the L dwarf companions.
Toward this end, both systems have been observed at multiple epochs starting
in May 1999 at the U. S. Naval Observatory in Flagstaff, Arizona.  Both Gliese
systems were observed with the 61-inch Strand Astrometric Reflector using the
ND9 camera.  This camera employs a 3mm ($\sim$40\arcsec) diameter attenuating
spot deposited upon an optically flat substrate that is mounted $\sim$1mm
in front of the Tek2K CCD.  This spot is located at the center of the 
$\sim$10\arcmin$\times$10\arcmin\ field of view and provides approximately
9.0 magnitudes of compensation.  Images were centered with G dwarf primaries
behind the spot so that both the bright primaries and L dwarf candidate
secondaries could be recorded simultaneously.  An astrometrically flat,
wide-I interference filter ($\lambda_{\rm c}\,\sim\,$8100 \AA;
FWHM$\,\sim\,$1910 \AA) was employed for the observations.

The astrometric observations are summarized in Table 1, which gives the epochs
of the observations, number of images taken, parallax factors in RA and Dec, and
the mean components of the projected tangents of the zenith distances for each
set of observations.  The parallax factors represent the effective triangulation
baseline (in AU) at the time of observation, and the sizable, non-monotonic
variation in parallax factor over the spread of epochs assures good separation of
parallax from proper motion.  The components of the projected tangent of the
zenith distance are relevant to potential contamination of the measures by
differential color refraction (DCR; cf. Monet et al.\ 1992) which will be addressed
below.

For the Gl 417 system, relative parallaxes and proper motions could be measured
for both the A and B components using 8-10 minute integrations.  A reference
frame of 16 stars, well distributed around both target objects, was initially
employed for the solution.  Table 2a summarizes the astrometric results for Gl 417,
with the first line presenting the HIPPARCOS (Perryman et al.\ 1997) results for
the A component for comparison.  The results for both A and B components from the
full USNO solution (13 exposures, 16 reference stars) are given in the second and
third lines.

In comparing the USNO parallax and proper motion results with the HIPPARCOS values,
it is important to remember that the USNO determinations are made from narrow-field,
differential measures and are, therefore, strictly ``relative'' to the mean parallaxes
and mean proper motions of the reference stars which are both set identically equal
to zero by the formal least squares reduction algorithms.  Relative parallaxes can
be ``corrected'' to absolute by estimating the mean parallax of the reference stars
employed, either from a galactic model, or better from (spectro)photometric
observations of the individual stars.  The HIPPARCOS results, on the other hand,
are derived from global solutions to whole-sky observations of many thousands of
stars and should, therefore, be truely ``absolute.'' As discussed below, photometric
observations for faint stars in the near proximity of such bright primary stars was
not generally successful.  However, based on other USNO parallax determinations
for fields at similar galactic latitudes as Gl 417 (+67.3\arcdeg) and where similarly
faint reference stars were observed individually with \hbox{V,V$-$I} photometry,
we can estimate a correction from relative to absolute parallax of approximately
0.9$\pm$0.2 milliarcsecond (mas) for the USNO determinations.  In making comparisons between the USNO
and HIPPARCOS parallax values, this amount should be added to the USNO relative
parallaxes presented in Table 2a.  When this is done, the agreement between the
USNO and HIPPARCOS parallax results are very satisfactory and justify adopting
the better determined HIPPARCOS astrometry for further discussions of the Gl 417
system.

Alternate, equally viable, solutions employing various subsets of the full USNO
data have been investigated.  For example, one of the observations at epoch 2000.35
is noticeably weak and the residuals for it stand out as $>$2$\sigma$ outliers.  If
that observation is rejected, the results presented in the fourth and fifth lines
of Table 2a are obtained.  In addition, 6 of the 16 reference stars employed in the
previous solutions are either quite weak on several frames or fall in the region of
the field affected by reflected light off of the compensation spot.  If we further
remove them from the astrometric solution, the results presented on the sixth and
seventh lines are obtained.  These are just example results from an array of
solutions carried out for various combinations of exposure frames and reference
stars.  They are presented to demonstrate that, although the observational data is
admittedly limited, the solutions they yield are indeed quite robust and do support
the physical relationship between the G dwarf primary and the L dwarf secondary --
that is, that the two objects are at the same distance and share a common proper motion.

The solutions presented thus far have not included any correction for differential
color refraction (DCR).  The wide-I filter employed for the observations has been
in routine use on the USNO parallax program since 1992 and a DCR correction for
it has been calibrated using the method described in Section 3.2 of Monet et
al. (1992).  The resulting relation is essentially the same as shown in figure 4 of
that paper for the STWIDER bandpass but with an amplitude only 0.252 times as
large.  As in the case of the STWIDER calibration, the wide-I calibration has only
been measured for stars as red as late M-dwarfs.  Extrapolation of this calibration
to redder L-dwarfs -- especially to late L-dwarfs, where the strong, broad KI
absorption alters the energy distribution within the wide-I bandpass significantly
-- is at best valid for observations confined to small range of hour angles.  As 
the data in Table 1 show, the astrometric exposures presented here were mostly
centered within 45 minutes of the meridian and there is only a small correlation
between projected tangent of the zenith distance and parallax factor.  Hence, we
anticipate that the effects of DCR on the astrometry presented should not be too
large.

Due to observational constraints, only rough estimates of the DCR effects are
possible.  Measures of V-I colors for both the L dwarfs and the astrometric 
reference stars were not possible using USNO telescopes and instrumentation.
The L dwarfs are simply too faint at V-band and even attempts to observe them
at R-band were unsuccessful.  Even the reference stars were, in most cases,
unobservable due to the larger amounts of scattered light from the G dwarf
primaries at V-band.  Hence, estimated colors have been adopted to illustrate
the potential effect that DCR might produce.

Examination of reference star photometry for 23 USNO fields at similarly high
galactic latitudes shows that the majority of reference stars have V-I colors
between 0.8 and 2.5.  Only an occassional star might be as blue as 0.6 or as
red as 2.8.  For six fields employing faint reference stars, the median V-I
colors of those stars ranged from 1.13 to 1.68.  Therefore, solutions including
DCR corrections were run assuming that all reference stars employed for the
Gl 417 field had either V-I=1.0 or V-I=1.7.  For Gl 417A the measured value
of V-I=0.67 was adopted from HIPPARCOS.  For Gl 417B,
a value of V-I=6.2 was assumed based on the measured value of V-I=6.2+/-0.5
obtained for 2MASSW 1507476$-$162738 which is bright enough to be measured at
V-band and which has a similar spectral type (L5) as Gl 417B (L4.5).

The astrometric solutions including these DCR estimates are given on the 
eighth through eleventh lines of Table 2a.  Small but significant changes in
the formal numbers are seen -- especially for the L-dwarf component -- when
these results are compared with the solutions neglecting DCR.  However, the
Table 2a results -- which include just a few examples from an array of
solutions carried out for various combinations of exposure frames, reference
stars, and DCR estimates -- demonstrate that, while the formal numbers can
change somewhat (as expected), in all cases the solutions support a physical
relationship between the G-dwarf primary (Gl 417A) and the L-dwarf secondary
(Gl 417B).  Hence, we feel well justified in adopting the better determined
HIPPARCOS astrometry in further discussions of the Gl 417 system.

The observational situation for the Gl 584 system was much more difficult. The
primary pair of G dwarfs, Gl 584AB, are recorded by the 61-in as partially
resolved; that is, not sufficiently separated for astrometric measure of
individual components and not sufficiently blended to permit reliable astrometry
of the combined light.  Furthermore, even when the AB pair was exposed to
saturation behind the 9.0 mag compensation spot, the L dwarf candidate companion
was not recorded due to its extreme faintness at I-band.  Following several
trial exposures with the AB pair moved to the edge of the CCD field or even off
the CCD, it was decided to leave the field positioned with the AB pair behind the
spot since this arrangement seemed to reduce the glare of scattered light from
the AB pair into the region of the L dwarf.  Exposure times of 20, 30, and 40
minutes were used at epoch 1999.38 and the results have relatively low S/N.
Single 90 minute exposures were used at epochs 2000.34 and 2000.51 and gave
better results.

Due to the increased amounts of scattered light from longer exposure times, the
faintness of Gl 584C, and the location of the target well into one quadrant of the
useable field, the results for Gl 584C are much less definitive.  Only 7 reference
stars were useable and 3 of these are very faint.  With only a limited number of
exposures and a poor reference frame, alternate solutions were not possible or
gave very unsatisfactory results.  Table 2b gives the HIPPARCOS solution for
Gl 584AB along with USNO solutions first without a DCR correction and then with
DCR estimates assuming a very uncertain V-I=6.5 for Gl 584C.  The agreement of
the non-DCR USNO parallax solution with the HIPPARCOS value appears to be highly
fortuitous, since the two DCR solutions show little agreement with the non-DCR
solution.  However,
the proper motions from all 3 USNO solutions agree well enough with the HIPPARCOS
results that it is very likely that Gl 584C has a physical relationship with
Gl 584AB and is at the same distance.  For further discussion we assume this to be
the case and adopt the HIPPARCOS astrometry for the Gl 584 system.

\subsection{Spectroscopic Confirmation}

Gl 417B and Gl 584C were observed in 1998 December and 1999 March
using the Low Resolution 
Imaging Spectrograph (LRIS; Oke et al.\ 1995) at the
10m W.\ M.\ Keck Observatory on Mauna Kea, Hawaii.  Table 3 contains a 
log of the 
observations\footnote{The other L dwarfs listed in this table are discussed in \S4.2}.
A 400 g/mm grating
blazed at 8500 \AA\ was used with a 1$\arcsec$ slit and 2048$\times$2048
CCD to produce 9-{\AA}-resolution spectra covering the range
6300 -- 10100 \AA.  The OG570 order-blocking filter was used
to eliminate second-order light.  The data were reduced and calibrated 
using standard IRAF routines.  A 1-second dark exposure was used to remove
the bias, and quartz-lamp flat-field exposures were used to normalize
the response of the detector. The individual stellar spectra were extracted
and sky subtracted
using the ``apextract'' routine in IRAF, allowing for the slight curvature of
a point-source spectrum viewed through the LRIS optics and using a template
where necessary. Wavelength calibration was achieved using neon+argon
arc lamp exposures taken after each program object. Finally,
the spectra were flux-calibrated using observations of 
standards LTT 9491, Hiltner 600, LTT 1020, and Feige 56 
from Hamuy et al.\ (1994).  As these data were taken as part of our general
spectral classification campaign, no data were acquired for telluric
absorption correction.  Hence, the earth's atmospheric
O$_2$ bands at 6867-7000, 7594-7685 \AA\ and H$_2$O bands at 7186-7273,
8161-8282, $\sim$8950-9300, $\sim$9300-9650 \AA\ are still present in
the spectra.  For all observations the slit was oriented at the parallactic
angle to eliminate problems associated with differential refraction.

As shown in Figure 2, the spectra of both objects confirm that these are L dwarfs.  The
classifications for these two objects, which are fully discussed in Kirkpatrick et al. (2000), are
L4.5 V for Gl 417B and L8 V for Gl 584C.
Both objects exhibit \ion{Li}{1} absorption at 6708 \AA, as illustrated in the insets in Figure 2. 
For both, the \ion{Li}{1} absorption has an equivalent width of 9 \AA, though the spectrum of Gl
584C in much noisier in this wavelength regime.  The presence of lithium absorption indicates that
these two L dwarfs are substellar (Rebolo et al., 1992).

\subsection{Spectrum Variability in Gl 584C}

A comparison of the 1998 Dec 24 and 1998 Dec 25 spectra of Gl 584C shows
evidence for variability in some of the feature strengths, prompting us to 
reobserve it again on 1999 Mar 04 and 1999 Mar 31.  A more detailed observing
log for Gl 584C is given in Table 4.  As shown in the table, each night
consisted of three or four back-to-back 20-minute integrations except for the
last observation on 1998 Dec 25, which was cut short by advancing twilight.

Each of these individual exposures is shown in Figure 3 except for those
of 1999 Mar 04, a night riddled by variable seeing and possibly light cirrus.  
There is excellent agreement
in the overall spectral shape of the eleven individual integrations, but
subtle differences are also seen.

Figure 4a shows details of the eleven spectra over the 8350-9050 \AA\ region.
There is no appreciable telluric absorption in this wavelength range, but the 
telluric emission, as shown in Figure 4b, is comprised of myriad OH and O$_2$
lines, all of which are blended at this resolution.  Despite the strong sky
lines, this telluric emission is
easily subtracted from the L dwarf spectra, leaving only slight residuals
at the 1\% level.  Individual 
spectra on 1998 Dec 25, for example, 
show excellent agreement despite the fact that the
O$_2$ and OH emission is greatly enhanced in the first, high-airmass spectra
relative to the later ones.   On any given night the shape of the spectrum in
this region repeats well, but discrepancies in the bandstrength 
near 8650 \AA\ exist from one night to the next.  Based on this figure we 
argue that this effect can not be a consequence of poor sky subtraction.  We
therefore conclude that the variability see in Figure 4a is inherent to the
L dwarf itself.

Figure 5 shows details of the same spectra but in the 9050-9750 \AA\ region.
Again, individual spectra for a particular night are internally 
self-consistent, but 
comparisons between nights reveal discrepancies.  In this case, the 
discrepancies are found in the 9250-9600 \AA\ H$_2$O absorption band, which
has (as stated in \S3.2) not been corrected for telluric H$_2$O absorption
at the same wavelengths.  Our experience at Keck on other clear runs where
telluric absorption corrections {\it were} applied in this wavelength regime
indicates that the depths of these telluric H$_2$O bands can vary by at least
10\% from one clear night to the next.  The change seen here between the 1998
Dec 24 and 1998 Dec 25 spectra is $\sim$15\%.  Hence, the changes seen in the
9250-9600 \AA\ region {\it may} simply be due to variations in telluric H$_2$O
absorption.  A rigorous monitoring program specifically designed to correct
for telluric absorption would be able to determine whether or not a portion of
the H$_2$O variations seen here is intrinsic to the L dwarf itself.

Figure 6a shows comparisons of the summed spectra for each night.  Overall,
the spectral slope repeats well from night to night as do the strengths of
the \ion{Cs}{1} lines.  Pairwise overplots in Figures 6b-f demonstrate the
nightly differences suggested by the individual spectra in Figures 4 and 5.
These overplots show the magnitude of the variability in the 8600-8700 \AA\ 
region longward of the CrH bandhead and the possible variability
in the H$_2$O band longward of 9300 \AA.

Variability has been reported previously in the spectra of other cool dwarfs.
Tinney \& Tolley (1999) reported variability in the integrated light of two 
30-\AA-wide bandpasses centered at 8570 and 8725 \AA\ in the M9 brown dwarf 
LP 944-20.  This is in the same spectral region where we see variability
in Gl 584C, and in the case of LP 944-20 the variation was detected in a 2-hour
time interval.  Variability has also been observed by Nakajima et al. (2000) 
in the near-infrared spectrum of the T dwarf SDSSp J162414.37+002915.6.  
Here the time span was only 80 minutes, but slight changes were nonetheless 
seen in the depths of some features assumed to be H$_2$O.  
Bailer-Jones \& Mundt (1999, 2001) have also detected photometric variability 
at $I$-band in three L dwarfs and one late-M dwarf.  In those studies, the
L1.5 dwarf 2MASSW J1145572+231730 was shown to have a variability whose
period does not repeat from one observing season to the next.

The variations seen in Gl 584C as well as the other cool dwarfs noted above
may be a consequence of inhomogeneities in the photosphere.  Such 
inhomogeneities would produce changes 
in the integrated light as the dwarf rotates.  This hypothesis is supported,
at least in part, by evidence that the rotation period of these cool dwarfs 
has a similar timescale as the observed variations, typically a few 
hours as measured via $v{\sin}i$ by Basri et al.\ (2000) and Schweitzer et
al.\ (2001).  What are these photospheric irregularities?  An analogy to 
low mass stars might suggest starspots -- regions of cooler temperature and 
increased magnetic activity.  An analogy to planets might suggest 
irregularly distributed ``clouds'' or bands in the atmosphere, the condensates 
blocking warmer layers below and producing chemical differentiation higher 
in the atmosphere.  Bailer-Jones \& Mundt (2001) find that photometric
variability is more common in L dwarfs than in late-M dwarfs; since 
chromospheric activity is seen to drop from late-M to early-L (Gizis et al.\
2000a), these authors suggest that starspots are unlikely to be the correct
explanation.  Moreover, the lack of significant periods in any of the
confirmed variables leads them to conclude that if the inhomogeneities are due
to surface features, they must evolve on timescales shorter than their 
observation baseline (typically 3-6 days).
Additional observations are badly needed to quantify the variations -- their
amplitudes, their timescales, and the spectral features most affected --
before we can begin to sort out the reasons behind them.

\section{Age and Mass Estimates}

Because the primaries in both systems are bright, well-studied G dwarfs,
we can use published observations to estimate their ages\footnote{For an
overview of age dating for main sequence stars of all types, see
Lachaume et al.\ (1999).}.  We can then use
those age estimates to constrain the masses of the L dwarfs themselves if we
assume that the G dwarf primaries are coeval with their
companions.

There are four main observational characteristics that can in principle 
be used to assign ages to solar-type stars:  (1) chromospheric and coronal
activity -- in general the older the star, the less active it 
is\footnote{A sustained level of activity is expected in very close
binary systems where tidal influences can counteract the trend of decreasing
rotation with time.  Such should not be the case for Gl 584AB since the pair
is a relatively wide system of separation 16 AU (Mason et al.\ 1999).  
Here, the observed measures 
of activity should still be a valid reflection of age.} ; (2)
metallicity -- this should be a direct reflection of the material, perhaps
processed and recycled in previous generations of stars, from which
it formed; (3) lithium abundance -- for
G dwarfs, the observed lithium abundance declines with time as the initial
lithium is slowly destroyed and not replenished; (4)
space velocity -- on average this increases with age as objects in the galaxy
undergo velocity boosts over their lifetimes due to encounters with giant
molecular clouds.

All four of these measurements
can also be intercompared on an ensemble of nearby stars to
identify young, co-moving, stellar kinematic groups.  Establishing
membership in a group is an even more valuable indicator of age since the
properties of all members of the group can be used to determine a more
reliable age for the group as a whole.  In \S4.1 we present the available
data to establish an age for each system.  In \S4.2 we use those age
determinations to estimate masses for the L dwarf companions.

\subsection{Ages}

\subsubsection{Estimates from Chromospheric and Coronal Activity}

\paragraph{X-ray luminosities:}

There are several measurements that can be made to determine the level of
chromospheric activity.  One such indicator is the fraction of
the star's luminosity emitted at X-ray wavelengths.  The logarithm of this
quantity ($\log (L_X/L_{bol})$, Table 5) is
plotted against $B-V$ color in Figure 7 for the G dwarfs Gl 417A, Gl 584AB, and the Sun.  Also shown
for comparison are the $\log (L_X/L_{bol})$ 
values for
members of the Pleiades and Hyades clusters as listed in Stern et al.\
(1995), Hempelmann et al.\ (1995), Micela et al.\ (1999), and 
Krishnamurthi et al.\ (1998).  The figure shows that both Gl 417A and
Gl 584AB are more X-ray active than the Sun with Gl 417A more active than
Gl 584AB.  Based on this plot alone, Gl 584AB looks older than the average
Hyad (625$\pm$50 Myr; Perryman et al.\ 1998).  Gl 417A appears somewhat
younger than the Hyades but not inconsistent with the age of the Pleiades
($\sim$125 Myr; Stauffer et al.\ 1998).

Gaidos (1998) derives a relation between $\log (L_X/L_{bol})$ and age 
for the evolving Sun using empirical relations for the time dependence of a
sunlike star's luminosity, the time dependence of rotation rate, and the 
dependence of X-ray luminosity on rotation period:
\begin{equation}
\log (L_X/L_{bol}) = -6.38 -2.6{\alpha}\log (\case{t}{4.6}) + \log [1 + 
0.4(1 - \case{t}{4.6})]
\end{equation}
where $t$ is age in Gyr and $\alpha$ is the exponent in equation (3) below.
Here equation (1) has been adjusted to yield the solar value 
listed in Table 5.  Because both Gl 417A and Gl 584AB are very similar in 
spectral type to the Sun, we can use this relation to estimate crude ages for 
the two G dwarf systems: 80-250 Myr for Gl 417A and $\sim$2 Gyr for Gl 584AB.

\paragraph{Emission in \ion{Ca}{2}:}

A second activity indicator is the fraction of the total luminosity emerging 
as chromospheric emission in the \ion{Ca}{2} H and K lines.  The logarithm of 
this quantity ($\log R^{\prime}_{HK}$, Table 5) is plotted against
$B-V$ color in Figure 8 for the G dwarf systems listed in Table 5.
Values for Hyades stars (Duncan et al.\ 1984) are also shown as well as mean
values for several other clusters.  Gl 417A emits a larger percentage of its 
total flux in
chromospheric emission than does Gl 584AB.  Whereas Gl 584AB has an
$R^{\prime}_{HK}$ index implying an age greater than the Hyades, Gl 417
is most likely of Hyades age or younger.  

Donahue (1993, 1998) gives a 
relation between age and chromospheric emission.  This relation is derived
from observed $\log R^{\prime}_{HK}$ values of stars in clusters spanning a
wide range of ages and is valid for ages older than 10 Myr:
\begin{equation}
\log (t) = 10.725 - 1.334R_5 + 0.4085{R_5}^2 - 0.0522{R_5}^3
\end{equation}
where $t$ = age in Gyr and $R_5 = (10^5)\log R^{\prime}_{HK}$.  Using this equation, we find an age estimate
for Gl 417A of $\sim$400 Myr and for Gl 584AB of $\sim$2.5 Gyr.

\paragraph{Rotational period:}

Rotational period can also be used as an age indicator since rotation rate
is coupled to the internal dynamo, which drives the level of chromospheric
and coronal
activity.  The rotational period ($P_{rot}$, Table 5) is plotted against
$B-V$ color in Figure 9.  The locus occupied
by Hyades and Pleiades stars is also shown (Stern et al.\ 1995; Hempelmann
et al.\ 1995; Krishnamurthi et al.\ 1998).  The rotation period of Gl 417A
implies that it is at least as young as the Hyades while that of Gl 584AB
implies an age older than the Hyades.  

The period of rotation is thought to increase as a power law over time:
\begin{equation}
P_{rot} \propto t^\alpha
\end{equation}
where the exponent most likely falls in the range $1/2 \lesssim \alpha \lesssim 1/e$ (Skumanich
1972; Walter \& Barry 1991).  If we take the 25.3-day rotation period of the Sun along with a solar
age of 4.6 Gyr, we can use equation (3) to estimate ages for the G dwarf systems of 150-400 Myr for
Gl 417A and 850 Myr-1.35 Gyr for Gl 584AB.

\paragraph{Chromospheric variations:}

Finally, the variation in chromospheric activity can also be used as a crude
chronometer of age for solar-type stars.  Stars with low-amplitude variability
generally have cyclical periods and are old stars like the Sun.  G dwarfs
of younger age ($\sim$1-2 Gyr) show moderate-amplitude variability and
occasional cyclical periods.  Much younger stars show larger variability
with no well defined periodicity.  As listed in Table 5, both Gl 417A and
Gl 584AB show variable periodicities and this implies, as do the other
activity indicators discussed above, that these G dwarfs are considerably
younger than the Sun.

\subsubsection{Estimates from Metallicity}

The values of [Fe/H] in Table 5 show that Gl 417A has solar metallicity
while Gl 584AB is more metal poor by 0.2 dex.  Figure 10 shows mean
[Fe/H] values for clusters and moving groups of known age, as measured from F and G stars by
Boesgaard (1989),
Boesgaard \& Friel (1990), and Friel \& Boesgaard (1992).  This figure demonstrates that there is little
correlation between metallicity and age, leading Boesgaard (1989) to the conclusion that the principal
factor determining a cluster's metal content is not its age but its position 
in the Galaxy.  Edvardsson et al.\ (1993, see their Fig.\ 31) show that the
age-metallicity relation for field stars also shows very large scatter.
Therefore, the metallicities of G dwarfs cannot be used as chronometers but may be suggestive of
membership in nearby clusters or moving groups.

\subsubsection{Estimates from Lithium Abundance}

In the study of stellar atmospheres, lithium is an important element because (1) it is easily fused
in stellar interiors and (2) it does not occur as a stable byproduct of any 
normal thermonuclear reaction (Herbig 1965; Bodenheimer 1965).
Hence, a star's current lithium content is just the fraction of its nascent lithium that it has not
yet destroyed.  Thus by assuming an intial lithium abundance and by measuring the current abundance,
the age of the star can be estimated.

However, the details of how a star recycles the material near its photosphere (i.e., those layers
visible to spectrographs) into deeper layers where lithium destruction occurs are complex and vary
as a function of mass.  K and early-M dwarfs, for example, are believed to be fully convective and
as a result of convective turnover will destroy all of their
primordial lithium.  For brown dwarfs below 0.06M$_\odot$, the interior never reaches temperatures
sufficient for lithium destruction, so these objects retain their full complement of primordial
lithium on timescales $<$100 Myr (Magazz{\`u} et al.\ 1993).  
In a mid-to-late F dwarf the convection zone is too shallow to allow
nuclear reactions to
destroy lithium at its base.  However, microscopic diffusion of Li atoms below the
convection zone will, slowly with time, deplete the star's lithium reserve.  For these stars, lithium
abundance is a good tracer of age (Boesgaard 1991).

The lithium abundance, $\log(N(Li))$, has been measured for mid-to-late F dwarfs in several well
studied clusters.  These mean values are plotted as a function of cluster age in Figure 11. 
Boesgaard (1991) represents the relationship by 
\begin{equation}
\log[N(Li)] = -0.30\log(t) + 5.33
\end{equation}
where $\log[N(Li)]$ is on the scale where $\log[N(H)] = 12.00$ and $t$ is the age of the F star in
Gyr.  In early type G dwarfs, like Gl 417A and Gl 584AB, the rate of lithium diffusion is higher
than for mid-to-late F stars, so the values of $\log[N(Li)]$ listed in Table 5 need to be adjusted
for use in equation (4).  In figure 1 of Boesgaard (1991), it is shown that the $\log[N(Li)]$ values
for stars of Hyades age decline by $\sim$0.3 dex from mid-F to G0.  Between mid-F and G1, the
difference is $\sim$0.6 dex.  Figure 9b of Soderblom et al. (1993c) shows that
for stars of Pleiades age, $\log[N(Li)]$ is roughly constant from mid-F to G1.
As further reference, the dependence of lithium abundance on age and on 
spectral type (T$_{eff}$) is especially well presented in Fig.\ 4 of Gaidos
et al.\ (2000).

We can therefore assume corrections to the lithium abundances for the G0 dwarf
Gl 417A and the $\sim$G1 dwarf pair Gl 584AB, plug those values into equation
(4), and determine for which values the resulting age is consistent with the
size of the correction assumed.  When we do this, we find that equation (4)
predicts
self-consistent ages of $\sim$600 Myr
for Gl 417A and $\sim$300 Myr for Gl 584AB.  
For late-F stars in a single cluster, however, star-to-star lithium abundances can vary
by $\pm$0.4 dex in $\log[N(Li)]$.  Moreover, as Figure 11 shows, the metal-rich
Hyades falls significantly above the mean relation suggested by the other
clusters, so the rate of lithium depletion may also have a dependence on
metallicity.  Another complication is that lithium depletion may also depend
on the rotational history of the star (Ventura et al.\ 1998).
Thus, these lithium-based
age estimates for Gl 417A and Gl 584AB should be 
considered as very crude estimates only.

\subsubsection{Estimates from Kinematics}

The $U,V,W$ space motions of Gl 417A and Gl 584AB place them squarely in the 
``young disk'' as defined by Eggen (1963) and
Leggett (1992).  (Note that those papers 
use the 
left-handed coordinate system -- i.e., where $U$ is defined as positive
{\it away} from the Galactic center -- as opposed to the right-handed
coordinate system employed here.)  Space motion is a poor indicator of age
for individual stars.  However, while old stars can have low space motions
relative to the Local Standard of Rest (LSR), very few young stars have
high velocities.  Thus, the observed motions (Table 5) are consistent with,
and even suggestive of, relatively young ages for both stellar systems
considered here.

\subsubsection{Estimates from Inclusion in Moving Groups}

The individual age estimates from above are listed in Table 6 for both of
the G dwarf primaries.  Using this information, we can determine whether
either is a likely member of a known moving group.

Gl 417A has been listed by Gaidos (1998) as a possible member of the Local
Association, which appears to comprise a kinematically coherent stream of
clusters and associations including $\alpha$ Persei and the Pleiades.  These
stars have ages substantially less than 300 Myr, most having ages of
20-150 Myr (Jeffries 1995).  The space motion of Gl 417A
(Table 5) is in reasonable agreement with the space motion\footnote{All
$U,V,W$ space velocities reported here are corrected to the LSR using a
solar motion of $U,V,W$ = +10.4, +14.8, +7.3 km/s, as described in Johnson
\& Soderblom (1987).}
 of stars thought
to belong to the Group, $U,V,W$ = -10,-21,-13 (Jeffries 1995), but the 
three-dimensional dispersion is a very large 9.9 km/s.  The metallicity is
also in good agreement with $\alpha$ Persei and the Pleiades (Figure 10).
It is clear, though,
that an age as young as 20 Myr for this star is not supported by the
evidence summarized in Table 6.  If this star is a member of the Local
Association, it would be one of the older members.  Based on the evidence
in Table 6 and the suggestion of Local Association membership, we assign
an age of 80-300 Myr to Gl 417A.

Gl 584AB has been listed previously as a possible member of
the Ursa Major Group, which has a traditional age of 300 Myr.  Although 
Gl 584AB is located on the sky near stars comprising the core of the
UMa Group and has a similar distance -- 18.6 pc for Gl 584AB and 21.7 pc for
the Group itself (Soderblom \& Mayor 1993a) -- the age estimates given above
and as summarized in Table 6 indicate that this G dwarf pair is significantly
older than 300 Myr.  Soderblom \& Mayor (1993a) arrive at the same conclusion
based on chromospheric activity comparisons alone.  We also note that the
metallicity of Gl 584AB, [Fe/H] = $-$0.20, is slightly lower than the
mean of the Group, [Fe/H] = $-$0.086$\pm$0.021, and significantly lower than
the [Fe/H] values for all eight UMa Group members with robust measures in
Boesgaard \& Friel (1990).  A comparison of the space motion of Gl 584AB with
the mean motion of stars forming the group nucleus, $U,V,W$ = +12.6,+2.1,$-$8.0
(Soderblom \& Mayor 1993a), also indicates that Gl 584AB is probably not a 
member since this star falls outside of the measured three-dimensional
dispersion of 1.7 km/s.  We use the individual age estimates in Table 6 to
derive an age range of 1-2.5 Gyr for this system, substantially older than the
age of the Ursa Major Group.

\subsection{Masses}

The spectral type of each L dwarf companion
can be used to get a crude estimate of temperature.  
Using these temperature estimates and the age estimates
from above, both companions can be placed on the theoretical H-R diagram and
masses estimated.  For
Gl 417B its spectral type of L4.5 V would place its temperature between 1600
and 1800 K based on the best current estimates (see, e.g., Kirkpatrick et al.
2000 and Basri et al. 2000).  For Gl 584C its spectral type of L8 V would 
place its temperature between 1300 and 1600 K, though a temperature nearer
1300 K is favored based on simple physical arguments (Kirkpatrick et al. 2000;
Reid et al. 2000).

Using the nongray atmospheres computed by Burrows et al. (1997), we can 
estimate masses for the two L dwarf companions.  As shown in Figure 12, this
gives $M=0.035{\pm}0.015 M_\odot$ for Gl 417B and $M=0.060{\pm}0.015 M_\odot$ 
for Gl 584C.  Note that the less massive object has an earlier spectral type
(is hotter) due to its much younger age.

\section{Comparison to Other L and T Dwarf Companions}

Several other L and T dwarf companions are also now known and can be compared
to Gl 417B and Gl 584C.  For three of these -- GJ 1048B, G 196-3B, and GJ 
1001B (LHS 102B) -- we have obtained far red spectra to place them on the same 
classification system as the other L dwarf companions.  Spectra of these
three objects were obtained with LRIS at Keck using the setup described in
\S3.2.  An observing journal for these three observations is given in the lower
portion of Table 3, and reduced spectra are shown in Figures 13 and 14.  
Using the classification scheme of Kirkpatrick et al. (1999b), we assign types 
of L1 V to GJ 1048B, L2 V to G 196-3B, and L5 V to GJ 1001B.  (Spectra of the
latter two objects are also shown in Mart{\'{\i}}n et al.\ 1999.)  We note that
GJ 1048B shows no \ion{Li}{1} 6708-\AA\ absorption (EW$<$1 \AA) or H$\alpha$
emission (EW$<$1 \AA).  G 196-3B shows \ion{Li}{1}
absorption with EW$\approx$6 \AA\ but no measurable H$\alpha$ emission
(EW$<$0.5 \AA).  GJ 1001B shows no \ion{Li}{1} absorption (EW$<$0.2 \AA) but
shows weak H$\alpha$ emission\footnote{This object is the latest type L dwarf
for which H$\alpha$ emission has been observed (see Kirkpatrick et al.\ 2000), 
thanks in part to the 
excellent signal-to-noise in this spectrum.} with EW$\approx$1.5 \AA.

Table 7a lists the physical data\footnote{We have used only
the age estimates for the primary stars to assign ages for each companion.  
As a result, our age estimates differ from some published values that 
incorporate comparisons between the properties of the companions and 
brown dwarf theoretical models.  Our approach allows us to use age
diagnostics deduced from the companion as independent tests of ages derived
from the primary alone.} for each of the L dwarf 
companions, and Table 7b lists data for the two known T dwarf companions.
Using our model-independent ages, we can plot these published companions
on the theoretical H-R diagram of Figure 12.  As with Gl 417B and Gl 584C,
temperatures for the L dwarfs of Table 7a have been estimated from spectral
types, with the temperature scales of Basri et al. (2000) and Kirkpatrick et
al. (2000) used to bracket the most likely ranges.  Temperature 
estimates for the T dwarf companions are taken from the literature.

Figure 12 shows that these L and T dwarf companions span a mass range from
$\sim$0.035 to $\sim$0.075 $M_\odot$.  The final mass estimates are given
in Tables 7a and 7b.  It is interesting to note that the dwarfs G 196-3B,
Gl 417B, and Gl 229B have nearly equal masses and thus to first order can be
thought of as snapshots of a 0.035-$M_\odot$ brown dwarf at three different
stages during its cooling history.  For an age even younger than that of the
Pleiades, such an object would be a late M dwarf.

For objects as cool as L dwarfs, the presence of lithium indicates a mass 
below $\sim$0.060 $M_\odot$.  Although we have not yet used the presence or
absence of lithium to constrain the ages or masses for the companions above,
we see that the results are self-consistent.  That is, for L dwarf companions
without lithium, all have independent mass estimates placing them above
$\sim$0.060 $M_\odot$.  Except for Gl 584C, all of the L dwarfs with lithium
have mass estimates placing them below $\sim$0.060 $M_\odot$.  The lithium
detection in Gl 584C favors an age in the younger part of our estimated age
range and a mass in the lower half of our mass range.

\section{Conclusions}

We have presented astrometric evidence that the G dwarfs Gl 417A and Gl 584AB
both have very low luminosity, common proper motion companions.  Spectroscopy
shows that both are L dwarfs and both have strong lithium absorption lines,
confirming that they are brown dwarfs.  The cooler object, Gl 584C, is a very
late L dwarf exhibiting bandstrength variations in its spectrum, perhaps
a sign of changing ``weather'' in the brown dwarf atmosphere.  Because the
primaries of both systems
are previously well studied, many parameters such
as absolute magnitude, age, and metallicity can already be assigned
to the L dwarfs themselves.  The L4.5 dwarf Gl 417B has $M_J = 12.9$, is 
between 0.08 and 0.3 Gyr old, and has [Fe/H] = $-$0.01.  Theoretical cooling 
tracks suggest that it has a mass of 0.035$\pm$0.015 M$_\odot$.  The L8 dwarf
Gl 584C has $M_J = 15.0$, just 0.4 magnitudes brighter in $M_J$ than the
well known T dwarf Gl 229B.  Gl 584C is between 1.0 and 2.5 Gyr old,
has [Fe/H] = $-$0.20, and has an inferred mass of 0.060$\pm$0.015 M$_\odot$. 
Both L dwarfs lie over 1000 AU away from their primaries. 

In a related paper, Gizis et al.\ (2001) 
conclude that the dearth of brown dwarf companions -- seen at separations
less than 3 AU from main sequence stars -- does not continue at very wide
separations.  This conclusion is based on the two 2MASS discoveries above
and on the 2MASS discovery of the T dwarf companion Gl 570D (Burgasser et 
al.\ 2000).  Each of these companions was found serendipitously during 
the course of field searches for L and T dwarfs, and the statistics for
such  widely separated brown dwarf companions is still poor.  A dedicated 
2MASS search specifically looking for companions to nearby stars is currently 
underway in an effort to bolster the statistics.  First results from this
survey will be presented in Wilson et al.\ (2001).

\acknowledgements

JDK acknowledges the support of the Jet Propulsion
Laboratory, California Institute of Technology, which is operated under
contract with the National Aeronautics and Space Administration.  
We thank Joel Aycock, Wayne Wack, and Barbara Schaefer for the outstanding
job they did in obtaining the Keck LRIS service observations on 31 Mar 1999;
Alice Monet and Steve Levine for obtaining some of the astrometric data
(Table 1) used for common proper motion verification;
John Stauffer and Dave Soderblom for enlightening discussions; and Lynne
Hillenbrand and John Carpenter for acquiring data in Dec 2000 used in the 
telluric absorption discussion of \S3.3.
INR and JL acknowledge funding through a NASA/JPL grant to 2MASS
Core Project science. The finder charts of Figure 1 make use of the Digitized Sky
Survey, which was produced at the Space Telescope Science Institute under
U.S.\ Government grant NAGW-2166.  Research in this paper has also relied
partly on photographic plates obtained at the Palomar
Observatory 48-inch Oschin Telescope for the Second Palomar
Observatory Sky Survey which was funded by the Eastman Kodak
Company, the National Geographic Society, the Samuel Oschin
Foundation, the Alfred Sloan Foundation, the National Science
Foundation grants AST84-08225, AST87-19465, AST90-23115 and
AST93-18984,  and the National Aeronautics and Space Administration 
grants NGL 05002140 and NAGW 1710.
This research has also made use of the SIMBAD database, operated
at CDS, Strasbourg, France.  This publication makes use of data from
the Two Micron All Sky Survey, which is a joint project of the University
of Massachusetts and the Infrared Processing and Analysis Center, funded
by the National Aeronautics and Space Administration and the National
Science Foundation.

\clearpage

\figcaption[kirkpatrick.fig1.ps]{Finder charts for both systems.  All four
images are 13.7 arcminutes on a side with north up and east to the left.
a) Left panels.
The top image shows the 1953-epoch POSS-I E-band image for Gl 417, and the 
bottom image shows the 1998-epoch 2MASS $K_s$ image.  The L dwarf companion, 
Gl 417B, is visible only in the near-infrared view.  b) Right panels.  The top
image shows the 1954-epoch POSS-I E-band image for Gl 584, and the bottom
image shows the 2000-epoch 2MASS $K_s$ image.  The L dwarf companion, Gl 584C,
is visible only in the near-infrared view.  
\label{fig1}}

\figcaption[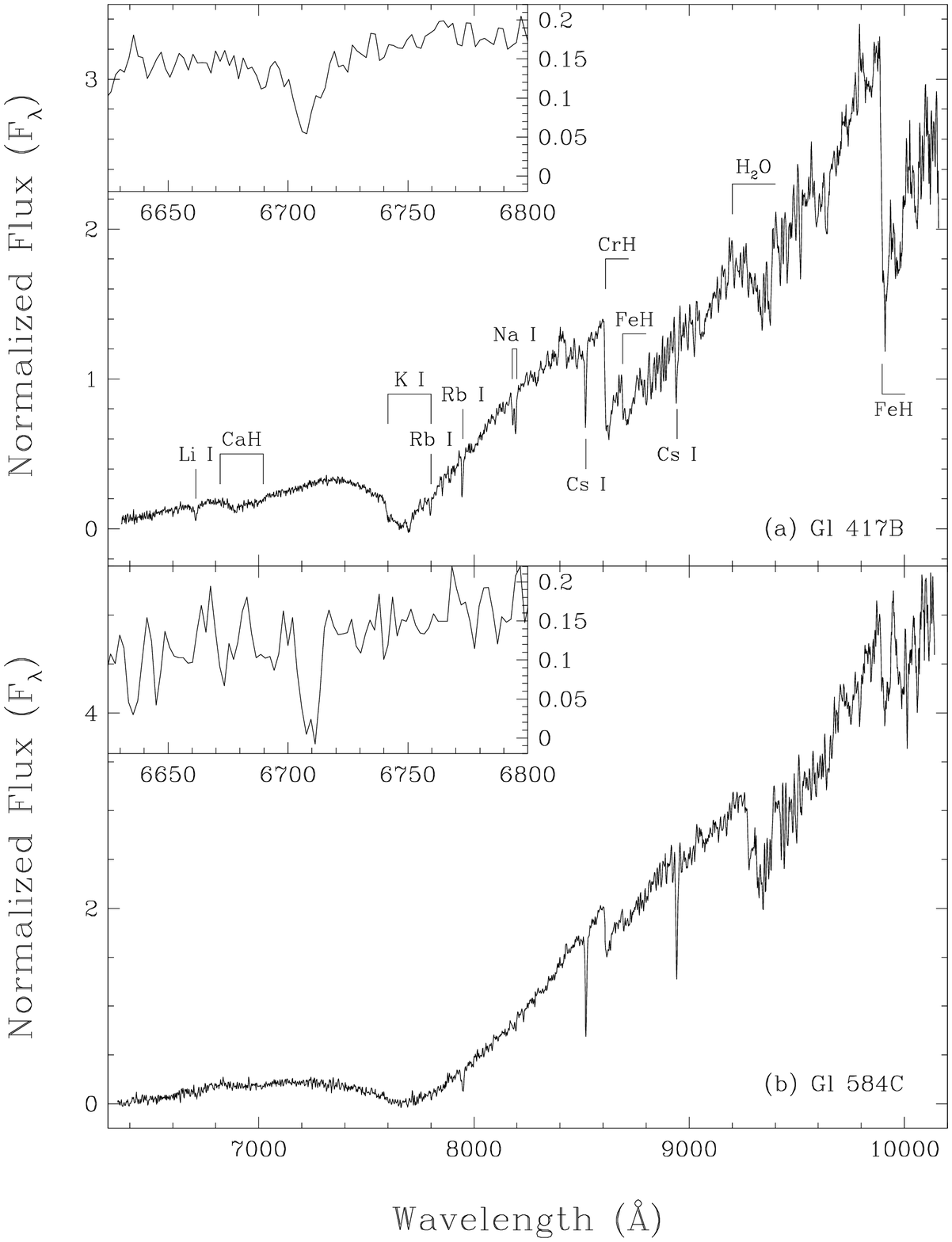]{Spectra of the L dwarfs. a) A 6,000-sec sum of the 
Gl 417B observations listed in Table 3.  Shown in the inset is a detailed 
view of the spectrum between 6600 and 6800 \AA\ to highlight the lithium 
absorption at 6708 \AA.  b) A 13,000-sec sum of the 1998 Dec 24, 1998 Dec
25, and 1999 Mar 31 observations listed in Table 3.  The inset shows a 
detailed view near the lithium absorption line at 6708 \AA, but here
only the 8,200-sec sum of the 1998 Dec data is shown as those data have the
best signal-to-noise in this region of the spectrum.  In both the full panels
and the insets, spectra have been normalized to one at 8250 \AA.  Prominent
spectral features have been labelled in panel (a).
\label{fig2}}

\figcaption[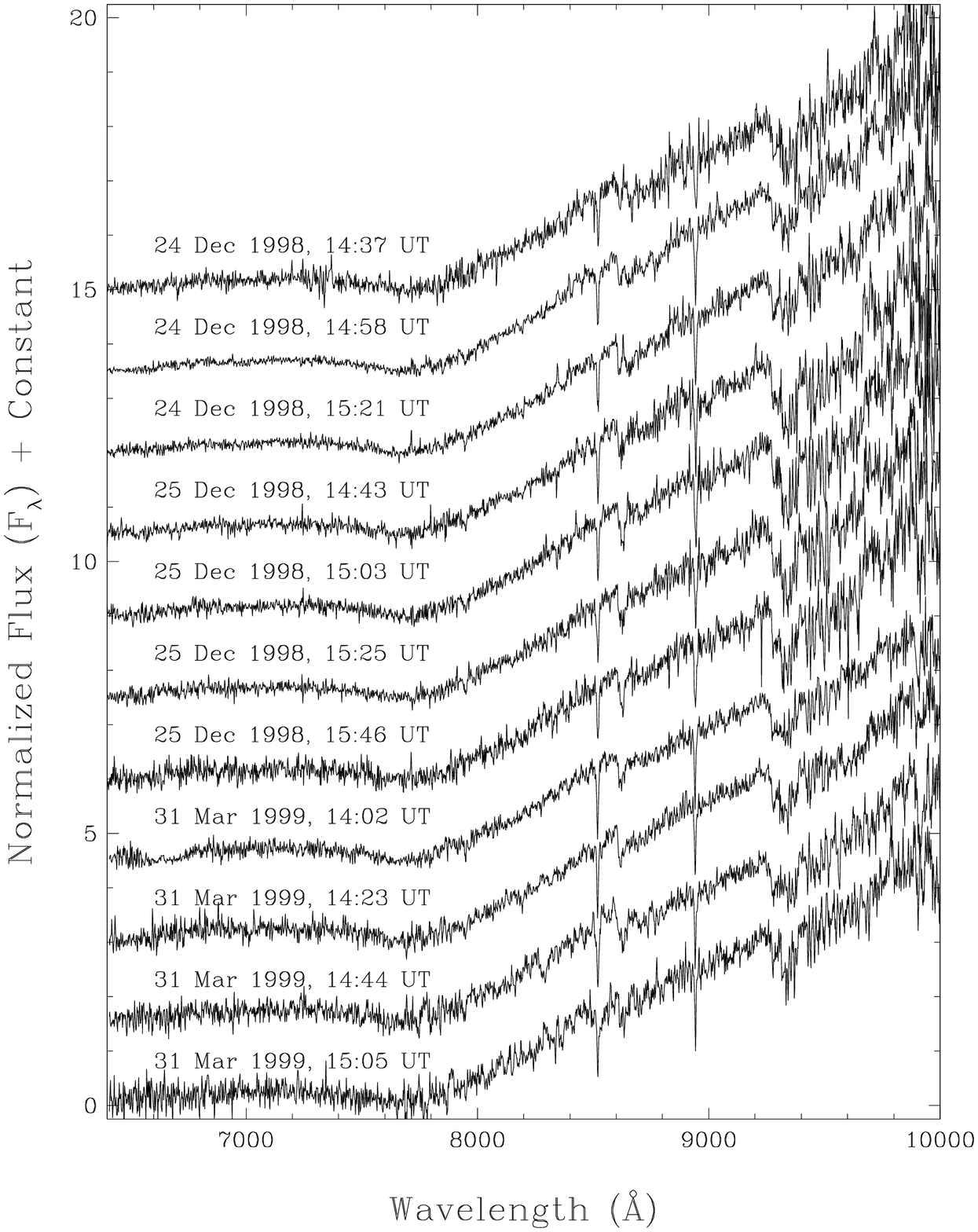]{Spectra of Gl 584C from each of the eleven individual 
integrations 
taken on 1998 Dec 24, 1998 Dec 25, and 1999 Mar 31 (UT).  Also listed are the
start times of each integration.  Each spectrum has been normalized to one at
8250 \AA, and integral offsets have been used to separate the spectra 
vertically.
\label{fig3}}

\figcaption[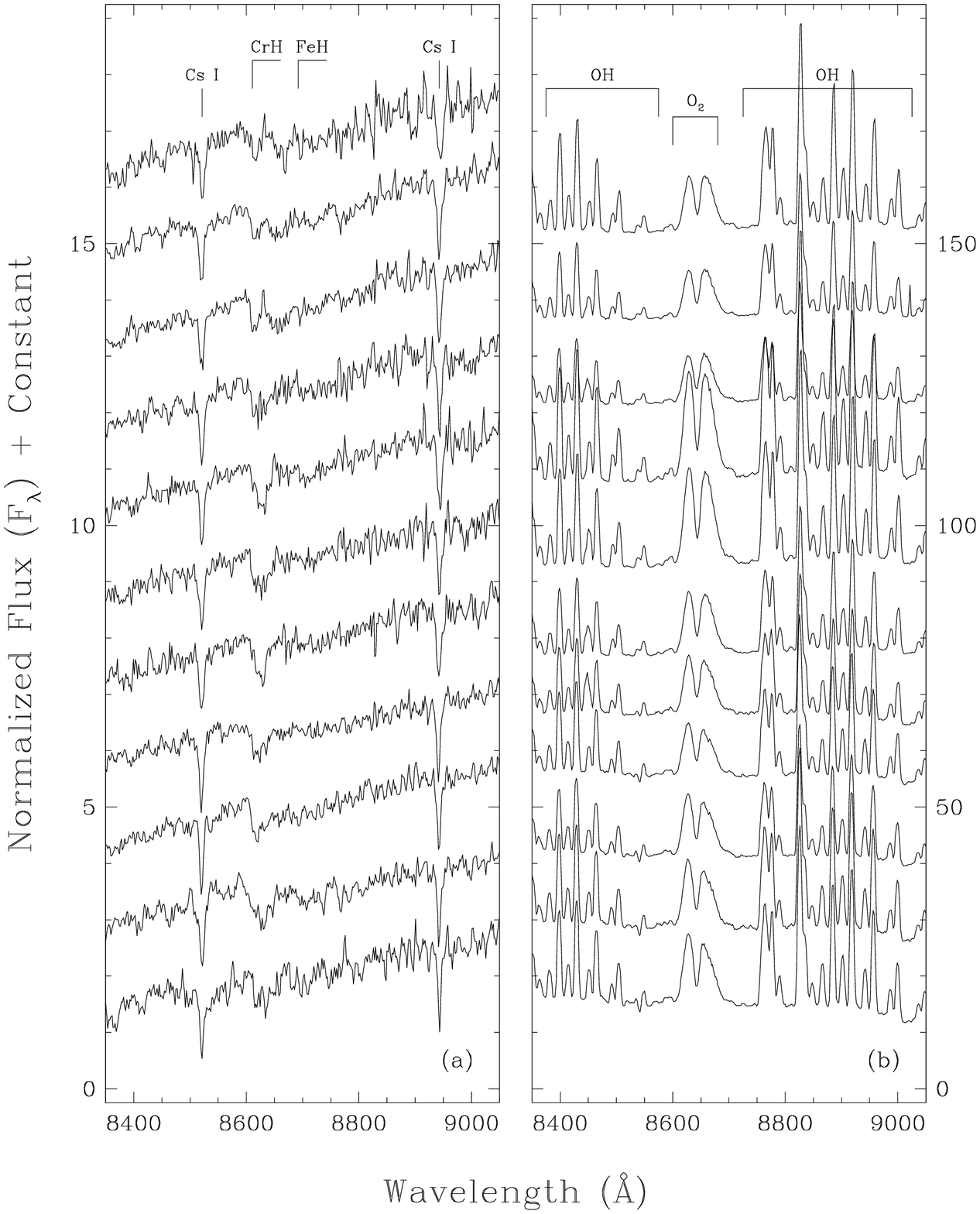]{a) Detailed views of the 8350-9050 \AA\ region for each
of the eleven spectra shown in Figure 3.  Noteworthy spectral features are 
marked.  b) Night sky emission spectra
for the same eleven integrations.  These spectra are a forest of OH and O$_2$
emission lines blended at this resolution.  The scaling and ordering of the
spectra are the same as in Figure 3.  
\label{fig4}}

\figcaption[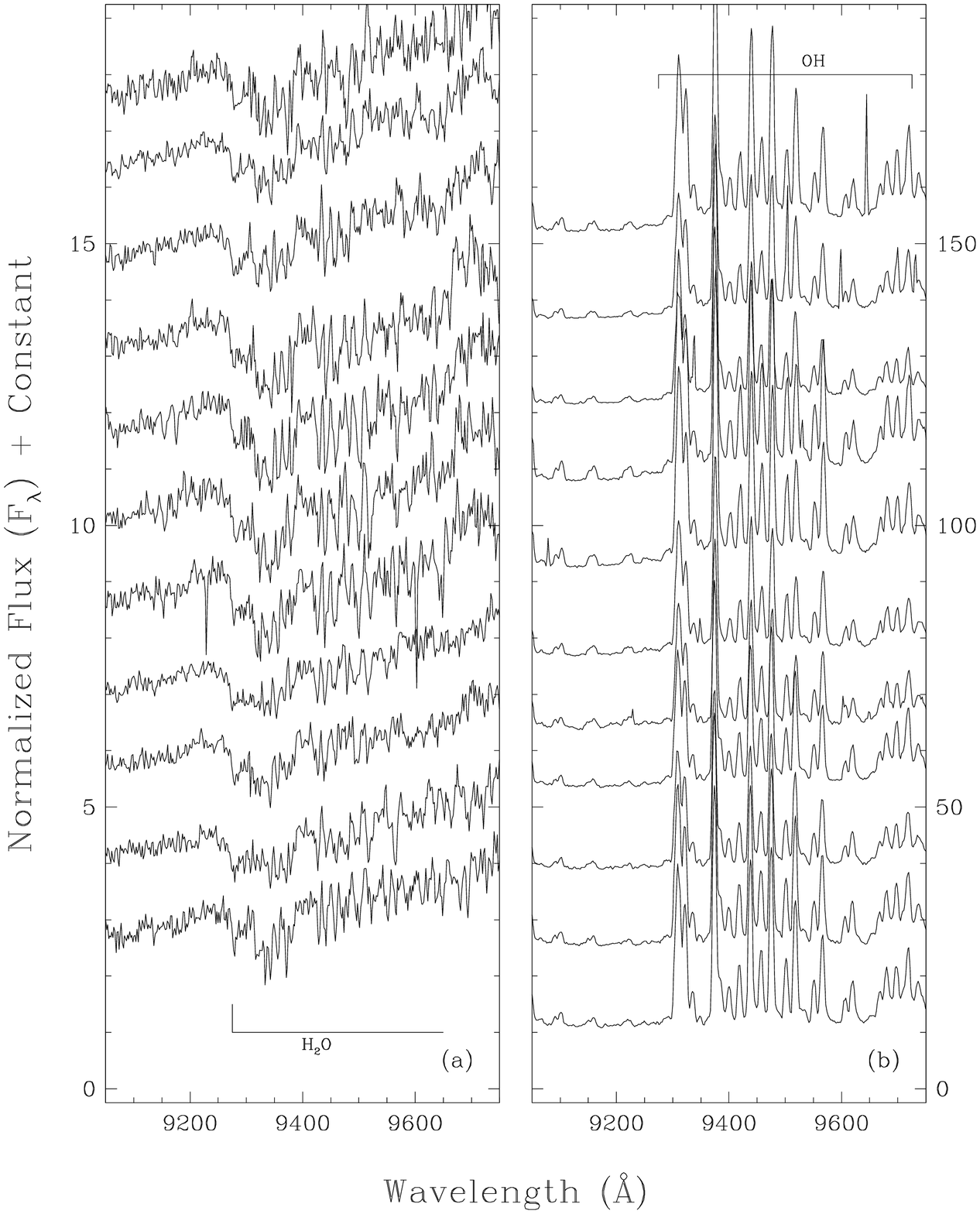]{The same as Figure 4 except for the 9050-9750 \AA\
region.\label{fig5}}

\figcaption[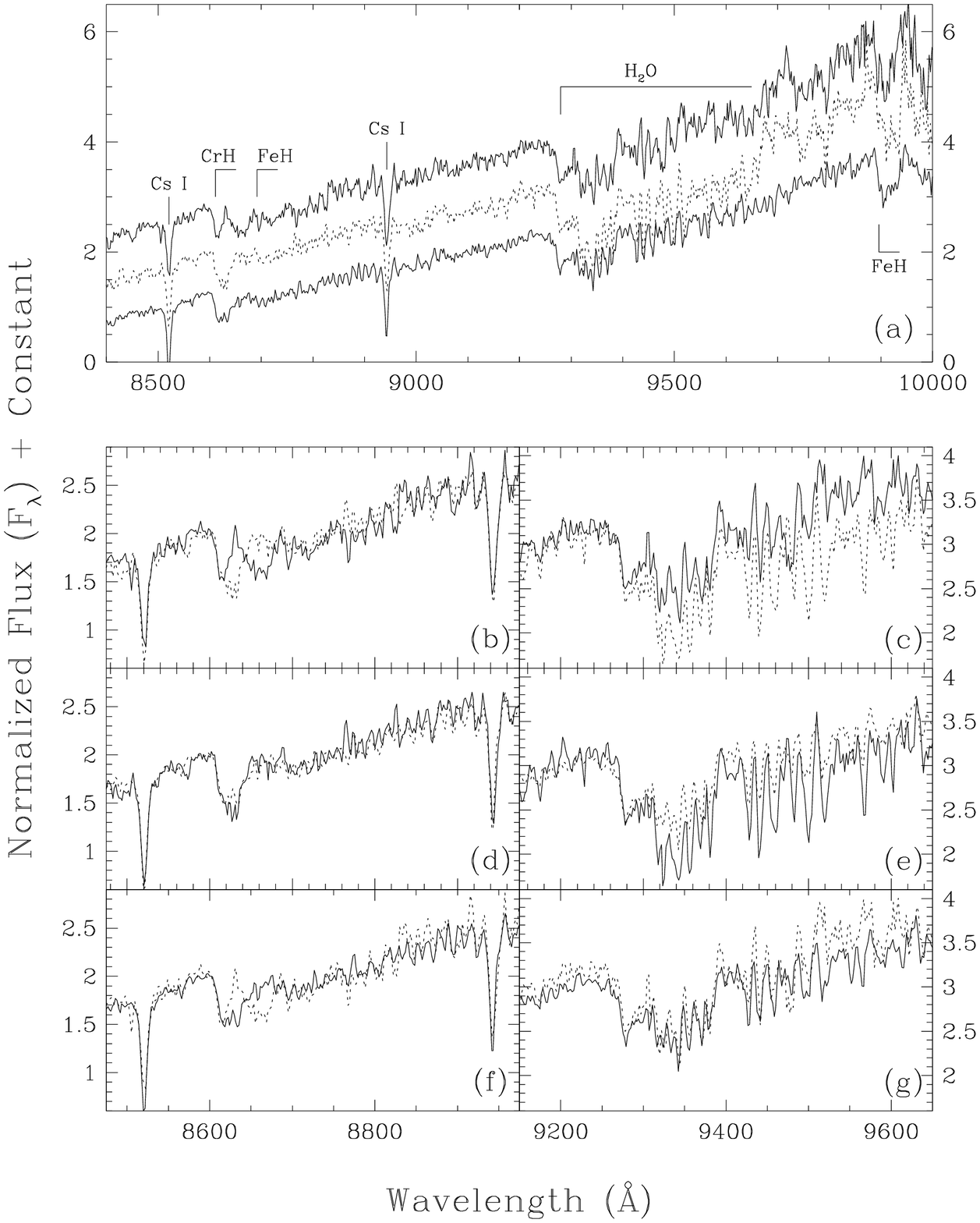]{Comparisons of the nightly spectra of Gl 584C.  a)
The region between 8400 and 10000 \AA.  The middle spectrum (dashed line)
is the summed spectrum taken on 1998 Dec 25 and normalized to one at
8250 \AA.  Above and below this are the summed spectra for 1998 Dec 24
(solid line, offset 0.75 upward) and 1999 Mar 31 (solid line, offset 0.75
downward).  The strongest spectral features are marked.  b,d,f) Pairwise
overplots of the spectra between 8475 and 8975 \AA.  c,e,g) Pairwise
overplots of the spectra between 9150 and 9650 \AA.  The pairs shown in
each panel are: b-c) 1998 Dec 24 (solid) and 1998 Dec 25 (dashed); d-e)
1998 Dec 25 (solid) and 1999 Mar 31 (dashed); f-g) 1999 Mar 31 (solid) and
1998 Dec 24 (dashed).
\label{fig6}}

\figcaption[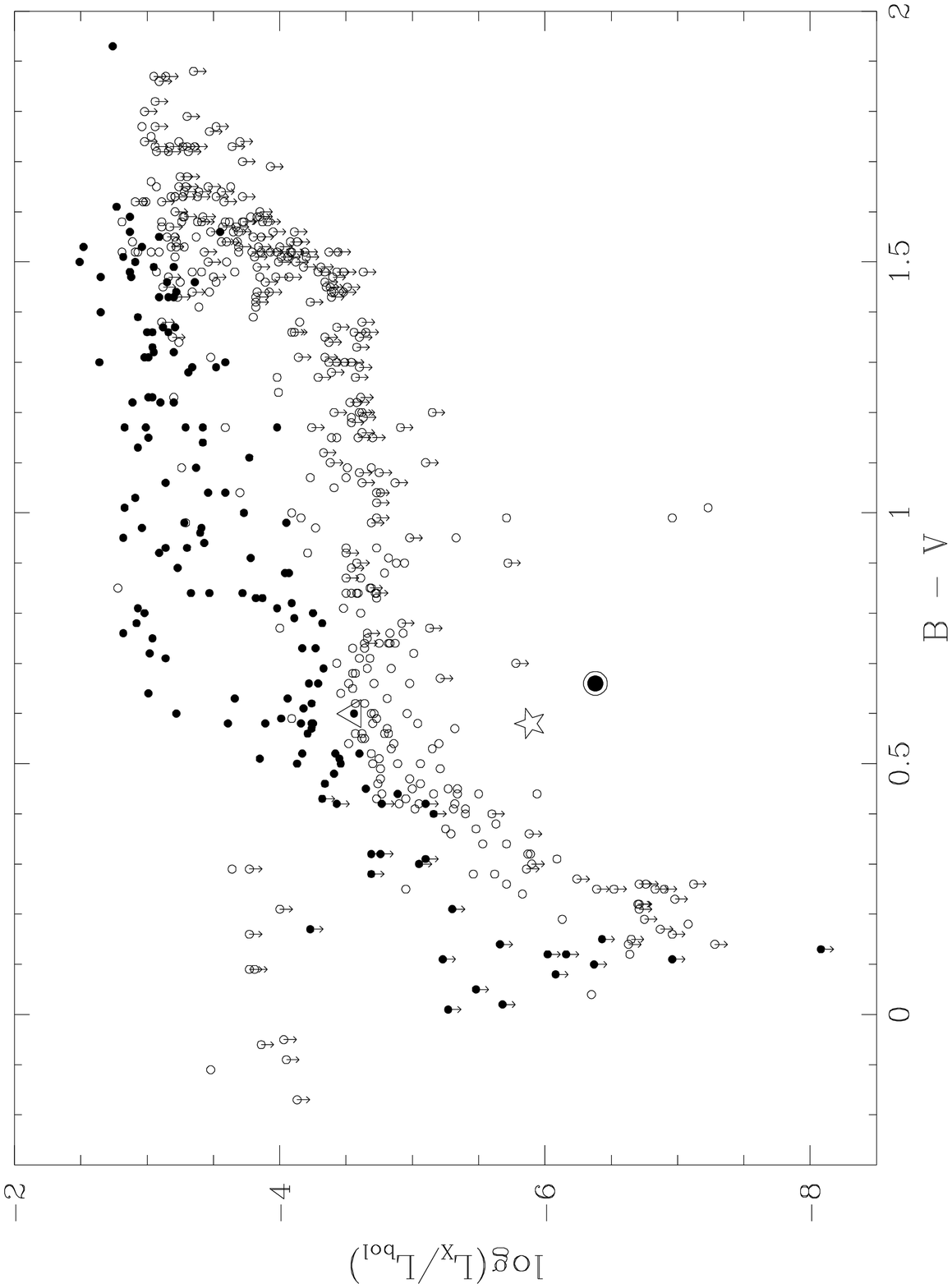]{$\log (L_X/L_{bol})$ vs.\ $B-V$.  Data for the Pleiades
(age $\sim$125 Myr) are shown by the small solid circles.  Data for the 
Hyades (age $\sim$625 Myr) are shown by the small open circles.  Upper limits
are denoted by downward arrows.  The position of the G dwarf Gl 417A is shown
by the large open triangle, and that of the G dwarf double Gl 584AB is shown by
the large open star.  The location of the Sun (age $\sim$4.6 Gyr) is shown by
the large encircled dot.  The sources for these data are given in the text.
\label{fig7}}

\figcaption[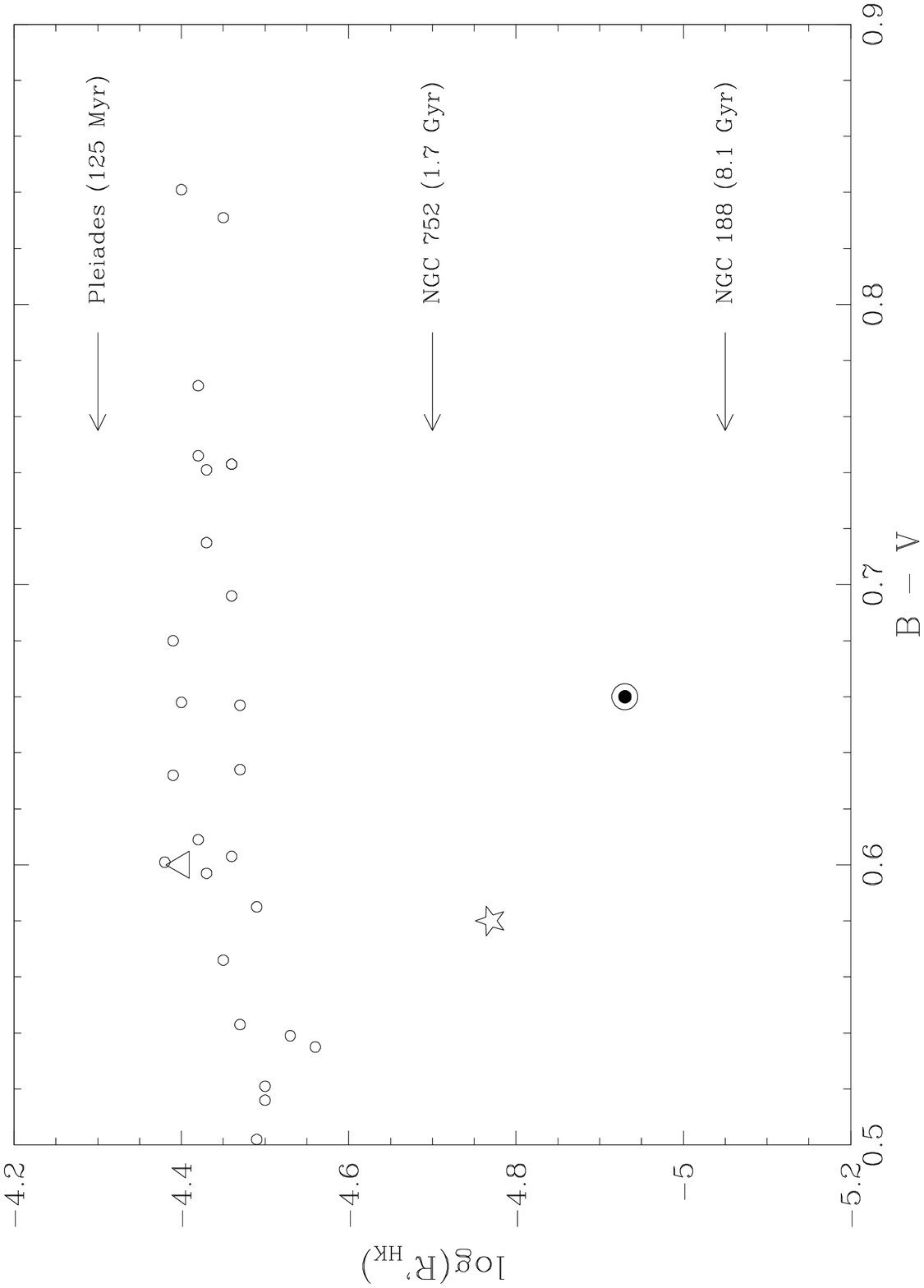]{$\log (R^{\prime}_{HK})$ vs.\ $B-V$.  Data for the Hyades
(age 625 Myr) are shown by the small open circles and are taken from Duncan
et al. (1984).  Also shown are mean $\log (R^{\prime}_{HK})$ values and ages 
for the Pleiades, NGC 752, and NGC 188 as taken from figure 1 of Donahue 
(1998).  The locations of Gl 417A, Gl 584AB, and the Sun are shown
using the same symbols as those in Fig.\ 7.
\label{fig8}}

\figcaption[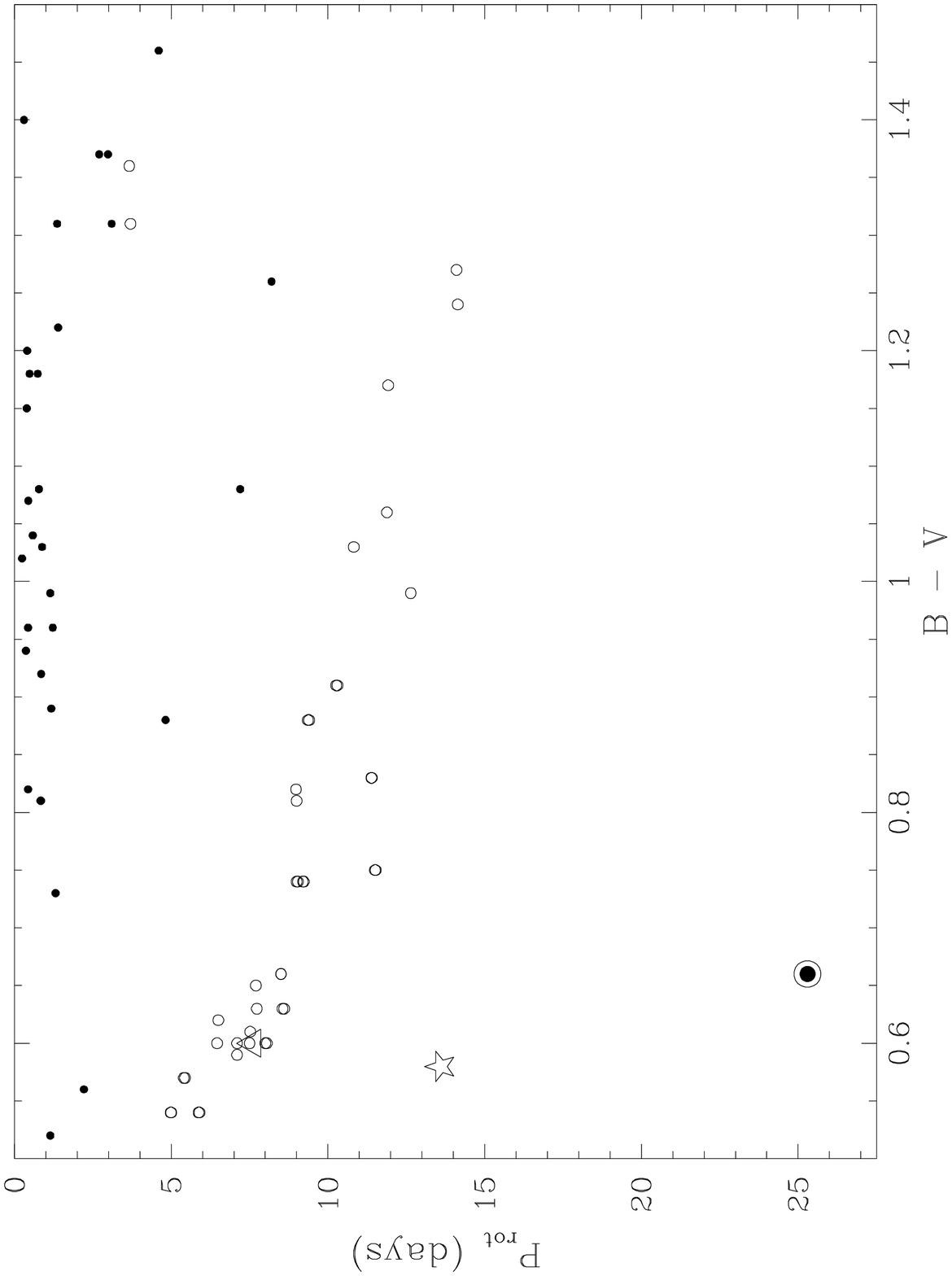]{$P_{rot}$ vs.\ $B-V$.  Symbols are the same as in Fig.\
7.  The sources for these data are listed in the text.
\label{fig9}}

\figcaption[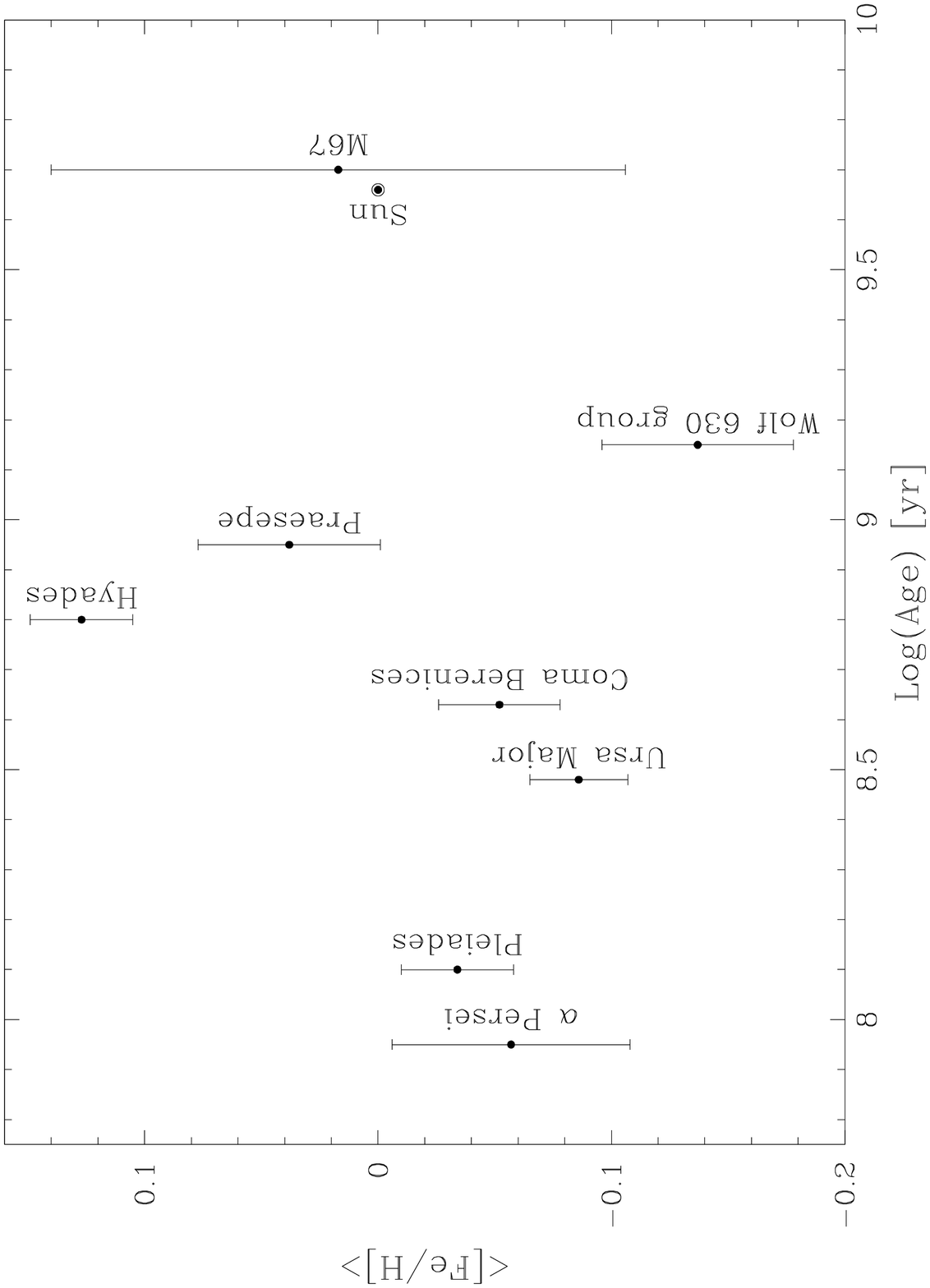]{Mean [Fe/H] vs.\ log age for various clusters and moving
groups.  Data are from Friel \& Boesgaard (1992). More recent age determinations
for $\alpha$ Persei (90 Myr, Stauffer et al. 1999), the Pleiades
(125 Myr, Stauffer et al. 1998), the Hyades (625 Myr, Perryman et al. 1998),
and Praesepe (900 Myr, Hambly et al. 1995) have been employed here.
\label{fig10}}

\figcaption[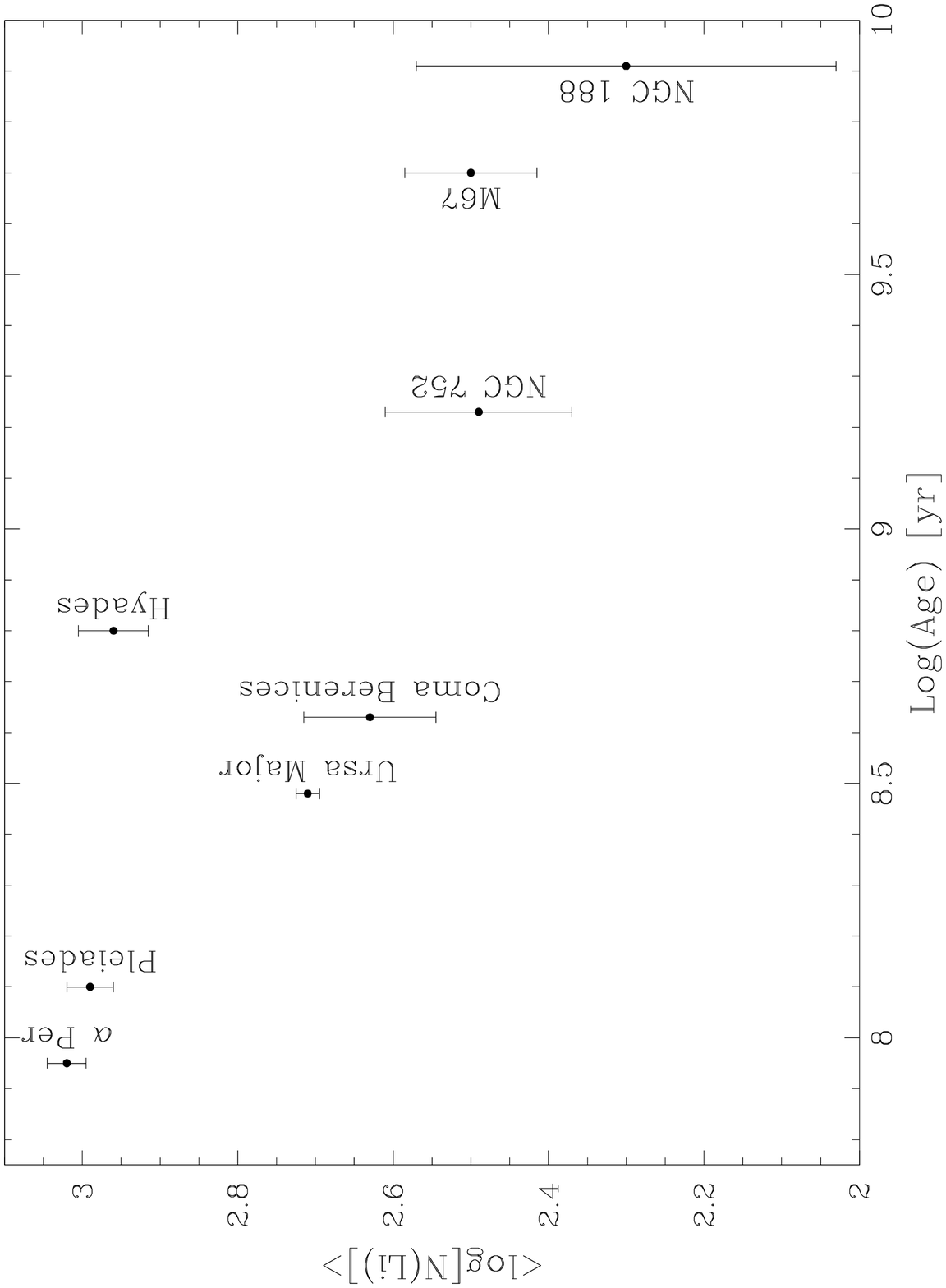]{Mean $\log[N(Li)]$ values for mid-to-late F dwarfs plotted
against age for various clusters
and moving groups.  Data are taken from Boesgaard (1991) but
supplemented with more recent age determinations as
detailed in the caption to Figure 10.
\label{fig11}}

\figcaption[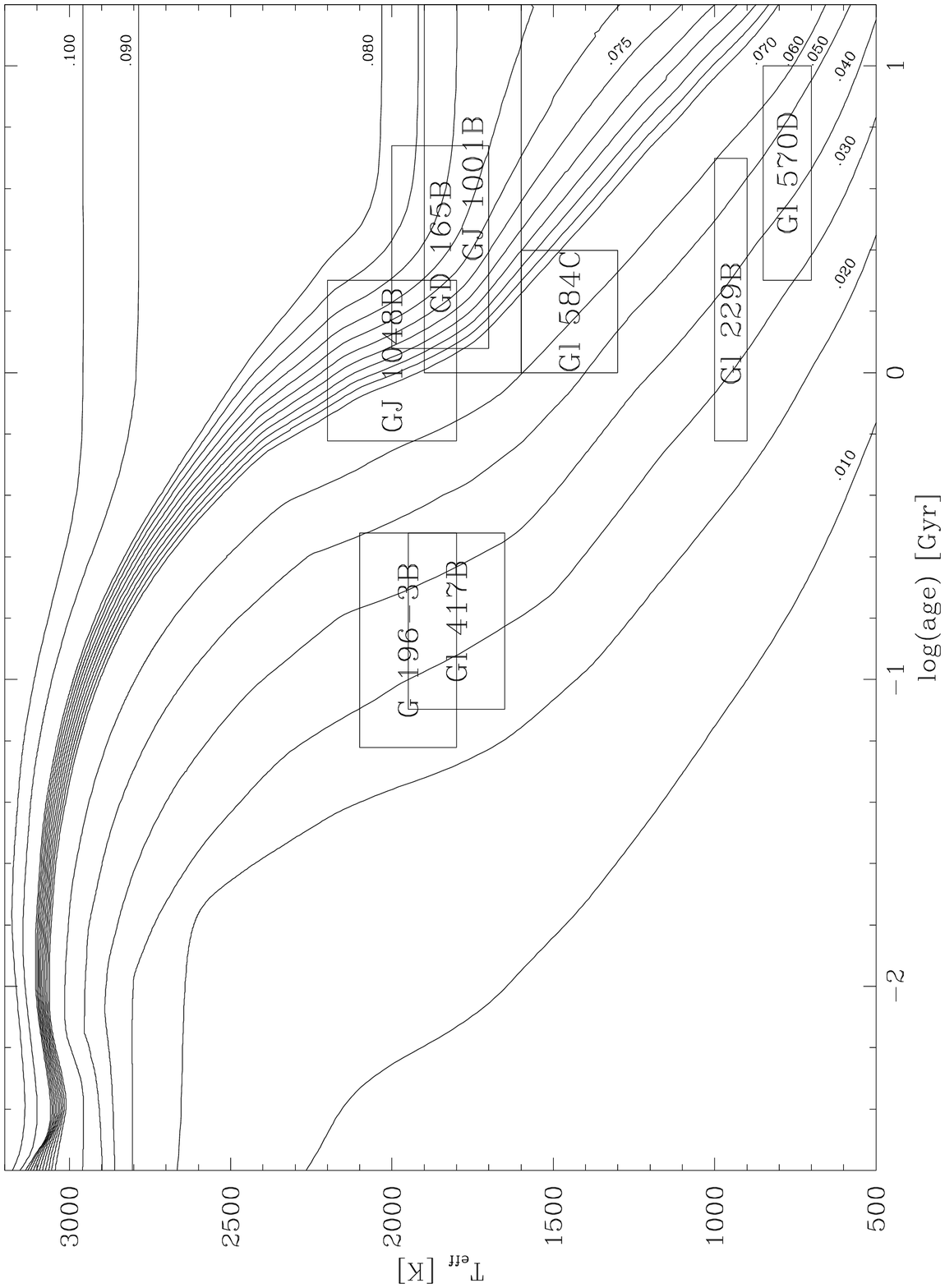]{Theoretical plots for low-mass stars and brown dwarfs
from Burrows et al.\ (1997).  The mass, in 
$M_\odot$, is shown along each track.  Positions are shown for each of
the L and T dwarf companions listed in Tables 7a and 7b.
\label{fig12}}

\figcaption[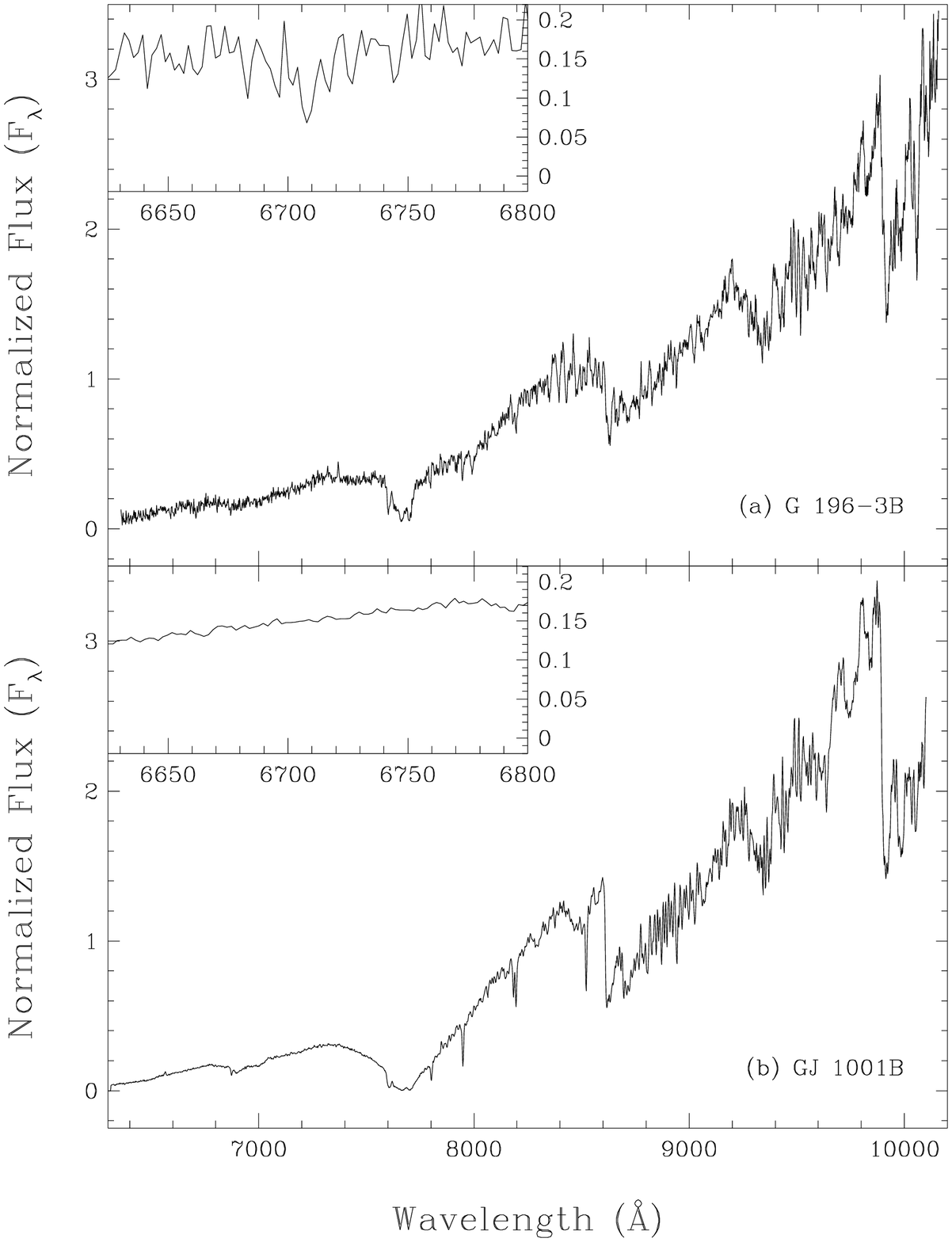]{Spectra of two other L dwarf companions. a) A 6,000-sec 
sum of the 
G 196-3B observations listed in Table 3.  Shown in the inset is a detailed 
view of the spectrum between 6600 and 6800 \AA\ to highlight the lithium 
absorption at 6708 \AA.  b) The 1,200-sec spectrum of GJ 1001B.  
The inset shows a detailed view near 6708 \AA\ demonstrating the lack of
lithium absorption.  In both the full panels
and the insets, spectra have been normalized to one at 8250 \AA.
\label{fig13}}

\figcaption[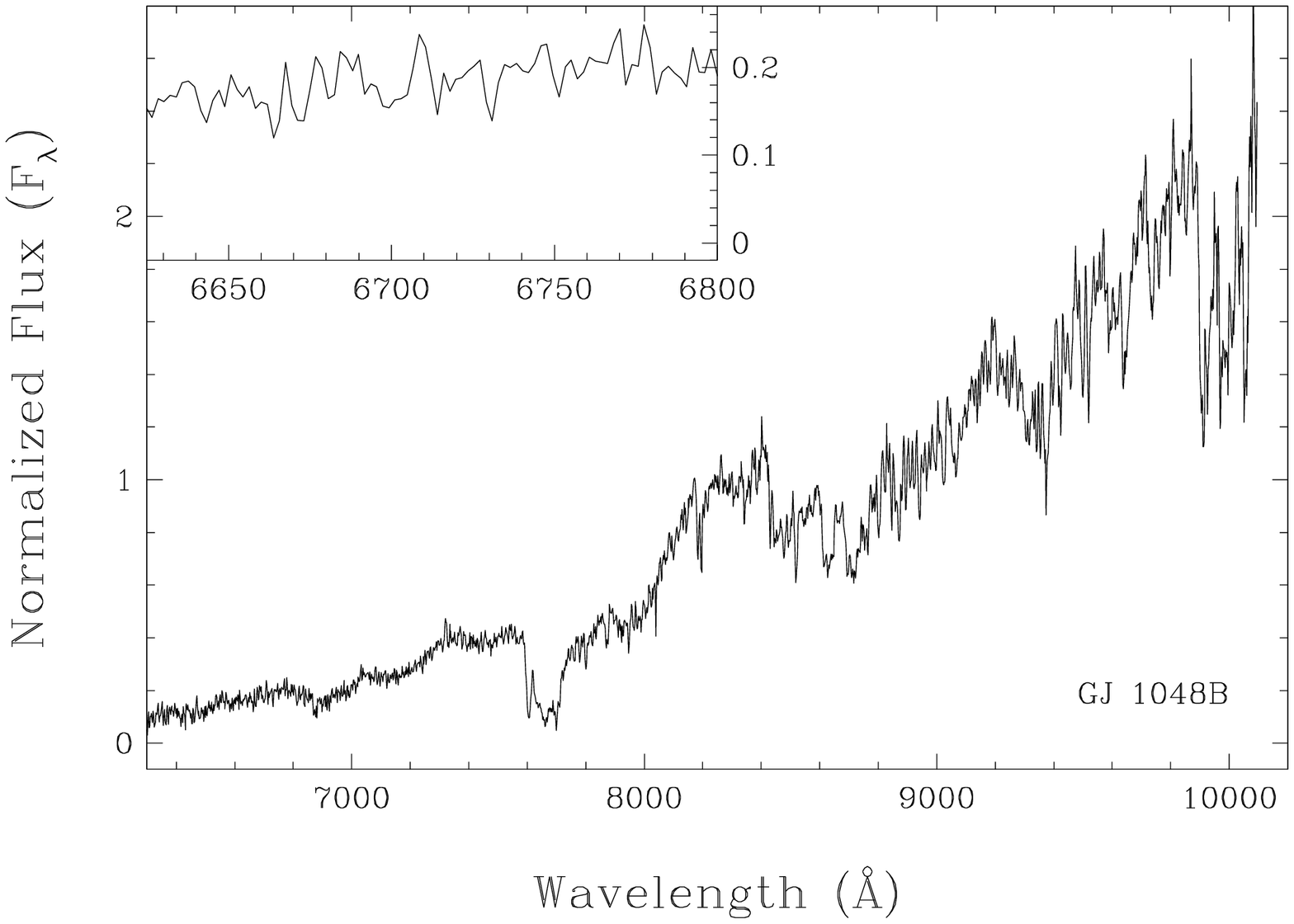]{Spectrum of another L dwarf companion, GJ 1048B,
as listed in Table 3.  As with Figures 2 and 13, the inset shows the
area between 6600 and 6800 \AA.  In the full panel
and the inset, the spectrum has been normalized to one at 8250 \AA.
\label{fig14}}

\vfill\eject

\begin{figure}
\figurenum{1}
\caption{The finder chart is too large for astro-ph submission. It can be
downloaded as a postscript file from 
http://spider.ipac.caltech.edu/staff/davy/papers/fig1.ps}
\end{figure}

\begin{figure}
\figurenum{2}
\plotone{kirkpatrick.fig2.ps}
\caption{}
\end{figure}

\begin{figure}
\figurenum{3}
\plotone{kirkpatrick.fig3.ps}
\caption{}
\end{figure}

\begin{figure}
\figurenum{4}
\plotone{kirkpatrick.fig4.ps}
\caption{}
\end{figure}

\begin{figure}
\figurenum{5}
\plotone{kirkpatrick.fig5.ps}
\caption{}
\end{figure}

\begin{figure}
\figurenum{6}
\plotone{kirkpatrick.fig6.ps}
\caption{}
\end{figure}

\begin{figure}
\figurenum{7}
\plotone{kirkpatrick.fig7.ps}
\caption{}
\end{figure}

\begin{figure}
\figurenum{8}
\plotone{kirkpatrick.fig8.ps}
\caption{}
\end{figure}

\begin{figure}
\figurenum{9}
\plotone{kirkpatrick.fig9.ps}
\caption{}
\end{figure}

\begin{figure}
\figurenum{10}
\plotone{kirkpatrick.fig10.ps}
\caption{}
\end{figure}

\begin{figure}
\figurenum{11}
\plotone{kirkpatrick.fig11.ps}
\caption{}
\end{figure}

\begin{figure}
\figurenum{12}
\plotone{kirkpatrick.fig12.ps}
\caption{}
\end{figure}

\begin{figure}
\figurenum{13}
\plotone{kirkpatrick.fig13.ps}
\caption{}
\end{figure}

\begin{figure}
\figurenum{14}
\plotone{kirkpatrick.fig14.ps}
\caption{}
\end{figure}

\end{document}